\documentclass[10pt,conference]{IEEEtran}

\usepackage{amsmath,amssymb,amsfonts,amsthm}
\usepackage{booktabs}
\usepackage{mathtools}
\usepackage{dblfloatfix}
\usepackage{pdfpages}
\usepackage[hidelinks]{hyperref}
\newtheorem{example}{Example}
\usepackage{caption}
\title{%
TENSKEL: A Combinatorial Observable Tensor for Structured Measurement and Reconstruction
}

\author{%
\IEEEauthorblockN{Yvan Richard}
\IEEEauthorblockA{
\href{https://orcid.org/0000-0003-4497-3843}
{ORCID: 0000-0003-4497-3843}
}
}

\begin{document}
\bstctlcite{IEEEexample:BSTcontrol}

\maketitle
\footnotetext{
An earlier version of this work was peer-reviewed and accepted for presentation at the IEEE ICIP 2026 Workshop on Quantum Computing and Quantum-Inspired Methods for Imaging (QCI 2026). It was withdrawn before publication in the conference proceedings because the author was unable to attend. This arXiv manuscript incorporates revisions based on the reviewers' comments.
}

\begin{abstract}
Many imaging problems seek to reconstruct underlying configurations from partial observable measurements. While reconstruction algorithms  operate on these measurements, the observable organization induced by the measurement process is rarely represented explicitly.

\vspace{0.25em}
\noindent
We introduce \textbf{TENSKEL}, a combinatorial observable framework for structured measurement and reconstruction based on tensor representations defined over discrete domains. Starting from a binary latent ensemble, the framework constructs a hierarchy of tensors coupling measurement contexts to a latent Pascal organization through successive aggregation and folding operations. Each measurement context induces an observable partition of the same latent ensemble, and the resulting tensor formulation makes explicit the associated combinatorial multiplicities, shell organization, degeneracies, and induced reconstruction geometry.

\vspace{0.25em}
\noindent
Rather than introducing a new reconstruction algorithm, this framework provides a mathematical representation of how latent configurations become organized under observation. The induced tensor kernel characterizes similarities between latent coordinates through their measurement-context responses, while regularized inversion provides a structured reconstruction of the latent representation from observable measurements. The binary construction further admits a natural multinomial extension to discrete simplex-supported latent representations.

\vspace{0.25em}
\noindent
Connections to Pascal cellular automata and structured discrete color mappings illustrate respectively compressed and multinomial realizations of the framework. More generally, TENSKEL provides a combinatorial basis for reasoning about the organization induced by observation, with potential relevance to computational imaging, inverse problems, structured sensing, and quantum-inspired measurement formulations.
\end{abstract}

\vspace{0.0mm}
\section{Introduction}
\noindent
Many imaging problems seek to reconstruct latent configurations from partial or aggregated measurements. Examples arise throughout computational imaging, compressed sensing, inverse problems, and related reconstruction tasks~\cite{donoho,candes}, where observable measurements provide only an incomplete description of the underlying configuration space.

\medskip
\noindent
Reconstruction from partial measurements is central to computational imaging, inverse problems, and discrete tomography~\cite{Kuba1999HistoricalOverview}. Although these fields differ in their sensing models and reconstruction algorithms, they share the common challenge of recovering latent configurations from aggregated observations. Existing approaches primarily address the consistency, uniqueness, and reconstruction of latent configurations, whereas the observable organization induced by the measurement process is rarely represented explicitly.

\medskip
\noindent
In this work, we introduce \textit{TENSKEL}, a combinatorial observable framework based on observable tensor operators and representations defined on discrete domains. Rather than proposing a new reconstruction algorithm, the framework provides a mathematical basis for representing and reasoning about the observable organization induced by measurement, from which structured reconstruction naturally follows.

\medskip
\noindent
The formulation considers a measurement-context domain
{\small
\[
\mathcal{C}
=
\mathcal{I}_{C}\sqcup\mathcal{O}_{C},
\]
}

\noindent
where $\mathcal{I}_{C}$ and $\mathcal{O}_{C}$ denote the inner and outer context regions, respectively, together with a latent simplex-supported region $\mathcal{I}_{L}$. The resulting tensor operator establishes an explicit relationship between measurement contexts and the latent Pascal organization, with aggregated measurements inducing context-dependent partitions and combinatorial degeneracies over the latent ensemble. Successive shell foldings then expose how this structure is organized across measurement-context and latent Pascal shells.

\medskip
\noindent
The framework is rooted in a classical combinatorial intuition: dice, Pascal structures, lattice paths~\cite{pascal,aigner_pascal,stanley,feller,graham_concrete}, and the transition from distinguishable microscopic configurations to aggregated statistical outcomes. In this sense, the tensor links two discrete organizations of the same ensemble: a uniform latent distribution over paths and an induced observable distribution over admissible counts~\cite{cover,feller}. The resulting observable multiplicities are therefore not imposed externally, but emerge from the combinatorial geometry of the fibres induced by these projections over the latent ensemble.

\medskip
\noindent
The proposed construction provides an explicit operator-theoretic bridge between distinguishable deterministic configurations and aggregated observable statistics. Pascal-type aggregation thereby becomes a structured measurement operator whose induced kernels, observable degeneracies, and inverse reconstruction problems can be studied directly. Although entirely combinatorial, the proposed observable formulation shares structural similarities with measurement-based formalisms encountered in quantum information and quantum-inspired imaging, thereby positioning observable tensor representations within a broader family of measurement-centered frameworks.

\medskip
\noindent
The binary-path formulation is also closely related to hypercube geometries, Hamming-weight organization, and error-correcting code structures~\cite{harary,hamming}. In this interpretation, observable shells correspond to combinatorial distance classes induced by aggregated path projections.

\subsection*{Notation}

\footnotesize
\setlength{\tabcolsep}{3pt}

\begin{tabular}{ll}
\toprule
Symbol & Meaning \\
\midrule
$\Omega_N$ & Binary latent ensemble $\{0,1\}^N$ \\
$\pi$ & Latent configuration / binary path \\
$\mathcal{C}$ & Measurement-context domain \\
$\mathcal{I}_{C}$ & Inner measurement-context region $p+q\le N$ \\
$\mathcal{O}_{C}$ & Outer measurement-context region $p+q>N$ \\
$\mathcal{I}_{L}$ & Latent Pascal simplex $a+b\le N$ \\
$M$ & Measurement-context cardinality $|\mathcal{C}|$ \\
$K$ & Latent simplex cardinality $|\mathcal{I}_{L}|$ \\
$\pi_{\mathrm{shell}}$ & Partition-indexing shell permutation \\
$(p,q)$ & Measurement-context coordinates \\
$S$ & Context shell $S=p+q$ \\
$a_p,b_q$ & Prefix-count measurements on $\pi$ \\
$(a,b)$ & Latent Pascal coordinates \\
$s$ & Latent Pascal shell $s=a+b$ \\
$\mathrm{Tab}(p,q,a,b)$ & Refined TENSKEL tensor \\
$T(p,q,s)$ & Latent-shell folded tensor \\
$T(S,s)$ & Context- and latent-shell folded tensor \\
$K_{\mathrm{Tab}}$ & Induced kernel $\mathrm{Tab}^{\top}\mathrm{Tab}$ \\
$C_{\mathrm{inner}}$ & Reconstructed latent simplex representation \\
\bottomrule
\end{tabular}

\normalsize
\setlength{\tabcolsep}{6pt}

%\clearpage
\vspace{2.0mm}
\section{Tensor Construction}

\subsection{Binary latent configurations and hypercube geometry}

\noindent
A latent configuration of length $N$ is represented by a binary path
{\small
\[
\pi=(\pi_1,\dots,\pi_N),
\qquad
\pi_j\in\{0,1\}.
\]
}

\begin{figure*}[!h]
\centering
\vspace{0mm}
\includegraphics[width=1.0\textwidth,trim=0 0 0 0,
clip]{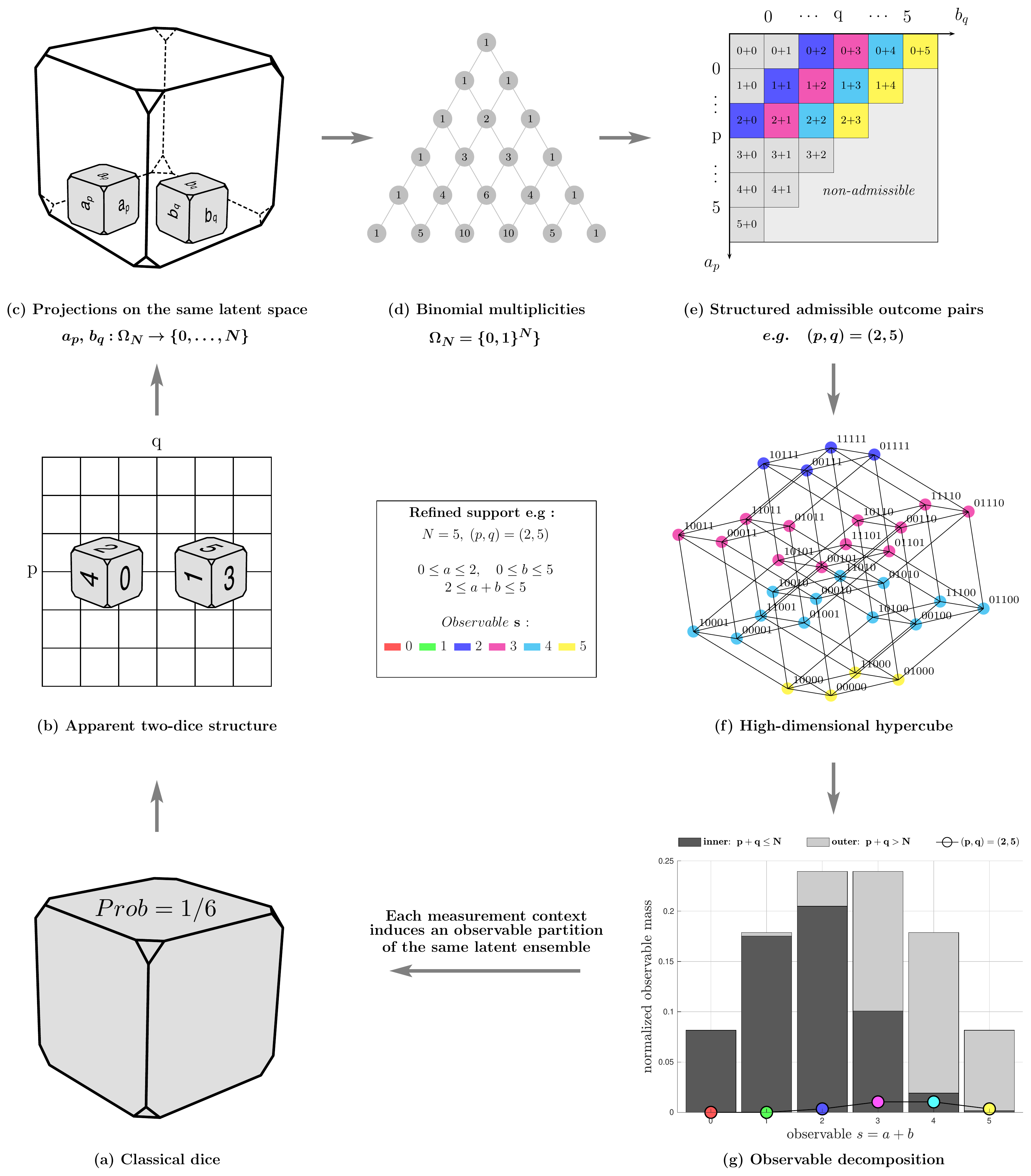}
\caption{
\footnotesize{
Conceptual organization of the TENSKEL observable construction for $N=5$.
(a) A classical die represents an elementary observable with uniform face probability.
(b) An apparent two-dice structure introduces two observable coordinates.
(c) The two coordinates are interpreted as complementary projections $a_p$ and $b_q$ acting on the same latent ensemble $\Omega_N=\{0,1\}^N$.
(d) Aggregation of the $2^N$ binary configurations produces Pascal/binomial multiplicities.
(e) For a fixed context $(p,q)$, only a constrained subset of observable pairs $(a,b)$ is admissible.
(f) The same latent hypercube supports the resulting context-dependent observable partition; the complete family of contexts is shown in Appendix~\ref{app:atlas}.
(g) Aggregation by $s=a+b$ yields the observable mass decomposition and separates contributions associated with the inner and outer measurement-context regions.
Together, the panels show how a fixed latent ensemble acquires different observable organizations under context selection, and how these organizations are encoded through admissibility and combinatorial multiplicity before normalization into probability distributions.
}}
\label{fig:qdice}
\end{figure*}

\noindent
Equivalently, the latent ensemble
{\small
\[
\Omega_N=\{0,1\}^N
\]
}

\noindent
is the vertex set of the $N$-dimensional binary hypercube. Each
configuration is observed through two overlapping prefix-count
measurements. The observable interpretation of these complementary
measurements, together with the induced admissible simplex geometry
and Pascal aggregation structure, is illustrated in
\textbf{Fig.~\ref{fig:qdice}}.
{\small
\[
\boxed{
\begin{aligned}
a_p(\pi)&=\sum_{j=1}^{p}\pi_j,
&&\text{prefix ones count},\\
b_q(\pi)&=\sum_{j=1}^{q}(1-\pi_j),
&&\text{prefix zeros count}.
\end{aligned}}
\]}

\noindent
Thus the tensor records how the binary hypercube appears under two
coupled prefix-count projections. The ordinary Hamming-weight map
partitions the hypercube into Pascal shells; here, the coupled prefix
projections refine this organization by inducing context-dependent
fibres over the latent ensemble, indexed by their outcomes $(a,b)$,
which are subsequently folded through $s=a+b$.

\medskip
\noindent
For $N=5$, Appendix~\ref{app:atlas} displays the complete family of
$(N+1)^2=36$ measurement contexts on the same latent hypercube. The
latent vertices remain fixed while the context $(p,q)$ changes the
induced observable partition, providing a direct visual realization
of the context-dependent organization encoded by the tensor.

\vspace{0.0mm}
\subsection{Refined observable tensor}

\noindent
Retaining both measurements defines the rank-4 observable tensor
{\small
\[
\mathrm{Tab}(p,q,a,b)
=
\bigl|
\{
\pi\in\Omega_N
\mid
a_p(\pi)=a,\;
b_q(\pi)=b
\}
\bigr|.
\]}

\noindent
Admissible states satisfy
{\small
\[
0\le a\le p,
\qquad
0\le b\le q,
\]
}

\noindent
together with the constraint
{\small
\[
\min(p,q)\le a+b\le \max(p,q).
\]}

\noindent
The closed form follows directly from the overlap geometry of the
two prefixes. Consider first $p\le q$ and divide a binary path into
three segments:
{\small
\[
\underbrace{1,\ldots,p}_{\mathrm{I}}
\quad
\underbrace{p+1,\ldots,q}_{\mathrm{II}}
\quad
\underbrace{q+1,\ldots,N}_{\mathrm{III}}.
\]
}

\noindent
Segment~I is shared by both measurements, segment~II belongs only to
the longer prefix, and segment~III is not resolved by either
measurement.

\medskip
\noindent
Fixing $a_p(\pi)=a$ selects $a$ ones among the first $p$ positions,
giving $\binom{p}{a}$ possibilities. If $x$ denotes the number of
ones in segment~II, then
{\small
\[
b
=
(p-a)+(q-p-x),
\]
}

\noindent
and therefore
{\small
\[
x=q-a-b.
\]
}

\noindent
The unresolved suffix of length $N-q$ contributes an independent
multiplicity factor $2^{N-q}$. Hence
{\small
\[
\mathrm{Tab}(p,q,a,b)
=
2^{N-q}
\binom{p}{a}
\binom{q-p}{q-a-b},
\qquad p\le q.
\]
}

\noindent
The case $p>q$ follows symmetrically. Splitting the path after
positions $q$ and $p$, fixing $b_q(\pi)=b$ selects $b$ zeros among
the first $q$ positions. If $y$ is the number of ones in the
non-overlapping segment $q+1,\ldots,p$, then
{\small
\[
a=(q-b)+y,
\qquad
y=a-q+b,
\]
}

\noindent
while the unresolved suffix contributes $2^{N-p}$.

\medskip
\noindent
Using
\(
\binom{n}{k}=0
\)
outside admissible ranges, both cases are therefore written as
{\small
\[
\boxed{
\begin{aligned}
\mathrm{Tab}(p,q,a,b)
=
\begin{cases}
\displaystyle
2^{N-q}
\binom{p}{a}
\binom{q-p}{q-a-b},
& p\le q,\\[10pt]
\displaystyle
2^{N-p}
\binom{q}{b}
\binom{p-q}{a-q+b},
& p>q.
\end{cases}
\end{aligned}}
\]
}

\noindent
The refined tensor therefore separates two sources of multiplicity:
combinatorial choices within the measured prefixes and unresolved
binary degrees of freedom outside the longest prefix.

\medskip
\noindent
Crucially, changing the measurement context does not alter the latent
ensemble $\Omega_N$; it alters the partition induced upon it. The tensor
therefore represents a family of observable organizations of the same
underlying configuration space.

\newpage
\vspace{0.0mm}
\subsection{Flat tensor view and inner--outer context structure}

\noindent
Using the shell permutation $\pi_{\mathrm{shell}}$, the measurement contexts $(p,q)$ are reordered according to the context shell
{\small
\[
S=p+q,
\]}

\noindent
while the latent Pascal coordinates $(a,b)$ are analogously ordered according to
{\small
\[
s=a+b.
\]}

\begin{figure*}[!h]
\vspace{-2mm}
\centering
\vspace{4mm}
\vspace{2.5mm}
\includegraphics[width=1.0\textwidth, trim=0 0 0 0,
clip]{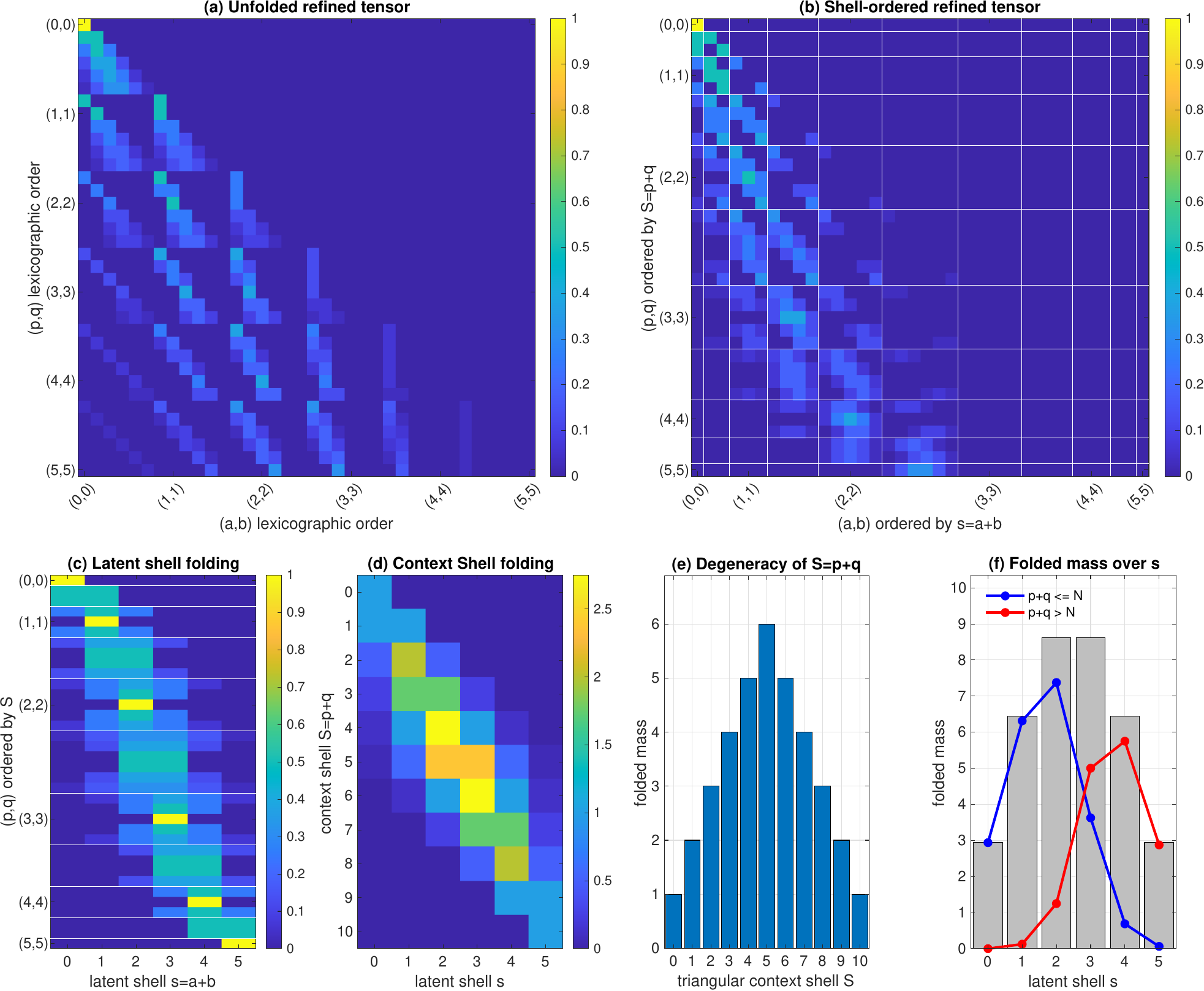}
\caption{
\footnotesize{
Tensor folding hierarchy of the refined TENSKEL construction for $N=5$. (a) Unfolded refined tensor in lexicographic ordering of the measurement-context coordinates $(p,q)$ and latent Pascal coordinates $(a,b)$. The tensor encodes the admissible couplings and combinatorial multiplicities induced between the two discrete domains. (b) Shell-ordered representation obtained by reindexing both domains according to the context shell
\(
S=p+q
\)
and the latent Pascal shell
\(
s=a+b.
\)
The permutation preserves the tensor entries while revealing a block geometry organized by the respective shell degeneracies. White separator lines indicate context- and latent-shell boundaries. (c) Latent-shell folding
\(
(a,b)\rightarrow s=a+b,
\)
which compresses the two-dimensional latent Pascal coordinates into the shell coordinate $s$ while retaining the individual measurement contexts $(p,q)$. The resulting representation $T(p,q,s)$ exposes the Pascal-type multiplicity structure induced by each measurement context. (d) Context-shell folding
\(
(p,q)\rightarrow S=p+q,
\)
applied to the representation in (c), yielding the compressed operator
\(
T(S,s).
\)
The resulting matrix couples context shells $S$ to latent Pascal shells $s$ and makes explicit the shell-to-shell organization induced by the successive tensor foldings. (e) Degeneracy of the context shell
\(
S=p+q.
\)
For $p,q\in\{0,\ldots,N\}$, the $(N+1)^2=36$ measurement contexts are distributed over $S\in\{0,\ldots,2N\}$ according to the classical triangular degeneracy of the square context lattice. (f) Total folded mass over the latent Pascal shell $s$, obtained from $T(S,s)$, together with the respective contributions from context shells satisfying
\(
S=p+q\le N
\)
and
\(
S=p+q>N.
\)
The decomposition shows how the latent-shell mass is distributed between the two regions of measurement-context space after the successive latent- and context-shell foldings.
}}
\label{fig:folding}
\end{figure*}

\noindent
The resulting shell-ordered tensor geometry and successive folding operations are shown in \textbf{Fig.~\ref{fig:folding}}.

\medskip
\noindent
The measurement-context domain
{\small
\[
\mathcal{C}
=
\{(p,q):0\le p,q\le N\}
\]}

\noindent
contains {\small $M=(N+1)^2$} contexts and separates naturally, with respect to the simplex boundary $S=N$, into the inner context region
{\small
\[
\mathcal{I}_{C}
=
\{(p,q)\in\mathcal{C}:p+q\le N\},
\]}

\noindent
and the outer context region
{\small
\[
\mathcal{O}_{C}
=
\{(p,q)\in\mathcal{C}:p+q>N\}.
\]}

\noindent
On the latent side, the admissible Pascal coordinates remain confined to the simplex-supported region
{\small
\[
\mathcal{I}_{L}
=
\{(a,b):a+b\le N\},
\]}

\noindent
whose cardinality is
{\small
\[
K
=
|\mathcal{I}_{L}|
=
\frac{(N+1)(N+2)}{2}.
\]}

\medskip
\noindent
Ordering the measurement-context axis according to $\mathcal{I}_{C}$ and $\mathcal{O}_{C}$, while the latent-coordinate axis is indexed by $\mathcal{I}_{L}$, gives the block decomposition
{\small
\[
\mathrm{Tab}
=
\begin{pmatrix}
T_{\mathrm{II}}\\
T_{\mathrm{OI}}
\end{pmatrix},
\]}

\noindent
where
{\small
\[
T_{\mathrm{II}}
\in
\mathbb{R}^{K\times K},
\qquad
T_{\mathrm{OI}}
\in
\mathbb{R}^{(M-K)\times K}.
\]}

\noindent
The two blocks respectively couple the inner and outer measurement-context regions to the same latent Pascal region:
{\small
\[
\mathcal{I}_{C}
\overset{T_{\mathrm{II}}}{\longrightarrow}
\mathcal{I}_{L},
\qquad
\mathcal{O}_{C}
\overset{T_{\mathrm{OI}}}{\longrightarrow}
\mathcal{I}_{L}.
\]}

\medskip
\noindent
The inner--outer decomposition reveals a marked asymmetry in the information carried by the two measurement-context regions. For the binary construction, numerical rank evaluation using scale-aware
singular-value tolerances, independently verified by modular rank computation, gives
{\small
\[
\boxed{
\operatorname{rank}(\mathrm{Tab})=K,
\qquad
\operatorname{rank}(T_{\mathrm{II}})<K,
\qquad
\operatorname{rank}(T_{\mathrm{OI}})<K.
}
\]
}

\noindent
Thus neither the inner nor the outer context region alone provides a complete representation of the $K$ latent Pascal coordinates. Their combination, however, restores full column rank in the complete tensor
$\mathrm{Tab}$. For the examined binary configurations, latent identifiability is therefore a property of the joint measurement-context organization rather than of either context region in isolation.

\medskip
\noindent
The corresponding rank structure is dimension-dependent under multinomial extension; its characterization beyond the binary case is left for further investigation.

\vspace{0.0mm}
\subsection{Observable folding hierarchy}

\noindent
The rank-3 tensor is obtained through diagonal aggregation:
{\small
\[
T(p,q,s)
=
\sum_{a+b=s}\mathrm{Tab}(p,q,a,b).
\]
}

\noindent
Diagonal aggregation groups the latent Pascal coordinates $(a,b)$ according to the shell variable
{\small
\[
s=a+b.
\]
}

\noindent
For a fixed measurement context $(p,q)$, this shell variable also indexes the aggregated observable outcomes induced on the latent ensemble.

\medskip
\noindent
The origin of the resulting Pascal structure can be seen directly from the overlap of the two prefixes. Defining
{\small
\[
m=\min(p,q),
\qquad
n=|p-q|,
\]
}

\noindent
exactly $n$ bit positions belong to the longer prefix but not to the shorter one. These are the only positions that can change the aggregated outcome $s$. The remaining $N-n$ binary degrees of freedom
do not change $s$ and contribute only multiplicity.

\medskip
\noindent
The number of contributing configurations is therefore determined by the number $s-m$ of selected contributions among these $n$ non-overlapping positions, giving
{\small
\[
\boxed{
\begin{aligned}
T(p,q,s)
&=
2^{N-n}
\binom{n}{s-m},\\
&\qquad
s\in\{m,m+1,\dots,m+n\}.
\end{aligned}}
\]
}

\noindent
Thus each context-dependent slice forms an $m$-shifted Pascal row of order $n=|p-q|$, scaled by the degeneracy factor $2^{N-n}$. The Pascal structure is therefore determined entirely by the
non-overlapping part of the measurement context, while the unresolved degrees of freedom determine its multiplicity.

\medskip
\noindent
The final context-shell folding is obtained through
{\small
\[
T(S,s)
=
\sum_{p+q=S}T(p,q,s).
\]
}

\begin{example}[$N=5$, $p=2$, $q=5$]
Here
\(
m=2
\)
and
\(
n=3
\),
so the three non-overlapping prefix positions determine the four admissible aggregated outcomes
\(
s\in\{2,3,4,5\}.
\)

\noindent
For fixed $(a,b)$,
{\small
\[
\mathrm{Tab}(2,5,a,b)
=
\binom{2}{a}
\binom{3}{5-a-b}.
\]
}

\noindent
Thus $T(2,5,s)$ is obtained by summing the entries along the diagonal strip
{\small
\[
a+b=s.
\]
}

{\small
\[
\begin{array}{c|cccccc}
 & b=0 & b=1 & b=2 & b=3 & b=4 & b=5\\
\hline
a=0 & 0 & 0 & 1 & 3 & 3 & 1\\
a=1 & 0 & 2 & 6 & 6 & 2 & 0\\
a=2 & 1 & 3 & 3 & 1 & 0 & 0
\end{array}
\]
}

\noindent
Reading the refined tensor by latent-shell diagonals gives
{\small
\[
\begin{array}{c|c|c}
s & a+b=s & T(2,5,s)\\
\hline
2 & 1+2+1 & 4\\
3 & 3+6+3 & 12\\
4 & 3+6+3 & 12\\
5 & 1+2+1 & 4
\end{array}
\]
}

\noindent
and therefore
{\small
\[
[
T(2,5,2),
T(2,5,3),
T(2,5,4),
T(2,5,5)
]
=
[4,12,12,4]
\]
}

\noindent
The context $(2,5)$ therefore exposes a shifted Pascal row of order $3$ inside the complete $2^5$-path ensemble. 
{\small
\[
[4,12,12,4]
=
2^2[1,3,3,1].
\]
}

\noindent
The factor $2^2$ records the two binary degrees of freedom that remain unresolved by the aggregated outcome. Thus observable aggregation reveals an embedded Pascal structure within the larger deterministic
path ensemble.
\end{example}

\begin{example}[$N=5$, $p=q=5$]
When the two prefix depths coincide, the non-overlapping region vanishes:
{\small
\[
n=|p-q|=0.
\]
}

\noindent
For every latent path,
{\small
\[
a_5(\pi)+b_5(\pi)=5,
\]
}

\noindent
so the aggregated observation contains a single outcome,
{\small
\[
s=5.
\]
}

\noindent
The refined latent Pascal coordinates nevertheless remain distributed along the diagonal
{\small
\[
(a,b)
\in
\{
(0,5),(1,4),(2,3),(3,2),(4,1),(5,0)
\},
\]
}

\noindent
with multiplicities
{\small
\[
1,\;5,\;10,\;10,\;5,\;1.
\]
}

\noindent
Consequently,
{\small
\[
T(5,5,5)
=
\sum_{a=0}^{5}\binom{5}{a}
=
32
=
2^5.
\]
}

\noindent
Thus all $32$ latent paths belong to a single aggregated observable class, while the refined tensor retains their internal Pascal organization. More generally, for every diagonal context $p=q$,
{\small
\[
s_{p,p}(\pi)=p
\qquad
\forall\,\pi\in\Omega_N,
\]
}

\noindent
which explains the single-class diagonal panels in the hypercube atlas of Appendix~\ref{app:atlas}.
\end{example}

\medskip
\noindent
The construction also inherits two exact combinatorial symmetries. Exchanging the two complementary prefix depths leaves the aggregated multiplicities unchanged:
{\small
\[
T(p,q,s)=T(q,p,s).
\]
}

\noindent
Moreover, under bitwise path complementation
$\pi\mapsto\bar{\pi}$,
{\small
\[
a_p(\bar{\pi})=p-a_p(\pi),
\qquad
b_q(\bar{\pi})=q-b_q(\pi),
\]
}

\noindent
so that
{\small
\[
s_{p,q}(\bar{\pi})
=
p+q-s_{p,q}(\pi).
\]
}

\noindent
Hence each context-dependent distribution satisfies the reflection symmetry
{\small
\[
\boxed{
T(p,q,s)
=
T(p,q,p+q-s).
}
\]
}

\noindent
The Pascal rows generated by the tensor are therefore symmetric about the midpoint $(p+q)/2$ of their admissible outcome interval.

\vspace{0mm}
\subsection{Normalization}

\noindent
Both tensors are normalized at fixed $(p,q)$:
{\small
\[
\sum_{a,b}\mathrm{Tab}(p,q,a,b)=2^N,
\qquad
\sum_s T(p,q,s)=2^N.
\]
}

\noindent
Every latent path therefore contributes exactly once to each measurement context; normalization changes the combinatorial counts into context-dependent probability distributions without changing their support or multiplicity structure.

\medskip
\noindent
The corresponding transition probability tensors become
{\small
\[
\boxed{
\begin{aligned}
\mathrm{Tab}_{\mathrm{prob}}(p,q,a,b)
&=
\frac{\mathrm{Tab}(p,q,a,b)}{2^N},
\\[3pt]
T_{\mathrm{prob}}(p,q,s)
&=
\frac{T(p,q,s)}{2^N}.
\end{aligned}
}
\]
}

\vspace{2mm}
\section{Observable Geometry and Reconstruction}

\noindent
The induced latent kernel
{\small
\[
K_{\mathrm{Tab}}
=
\mathrm{Tab}^{\top}\mathrm{Tab}
\]
}

\noindent
induces a similarity geometry over latent Pascal coordinates through the overlap of their measurement-context representations. Latent coordinates producing similar context-dependent response patterns therefore acquire non-trivial geometric proximity within the induced kernel space.

\medskip
\noindent
From an inverse-problem perspective, the tensor acts as a structured measurement operator linking the latent Pascal representation to measurements defined over the measurement-context domain.

\medskip
\noindent
This naturally leads to regularized reconstruction formulations of
the form
{\small
\[
C_{\mathrm{inner}}
=
(\mathrm{Tab}^{\top}\mathrm{Tab}+\lambda I)^{-1}
\mathrm{Tab}^{\top}Y,
\]
}

\noindent
where $Y$ denotes the observable measurements and $C_{\mathrm{inner}}$ the reconstructed latent representation.

\medskip
\noindent
Because observable aggregation is combinatorially organized, the inverse problem differs from generic linear reconstruction: the admissible latent domain already carries an intrinsic shell organization and simplex geometry inherited from the tensor
construction.

%\medskip
%\noindent
%The observable tensor therefore provides:
%\begin{itemize}
%\item a combinatorial aggregation operator,
%\item a structured observable geometry,
%\item an induced kernel space,
%\item and, through regularized inversion, a reconstruction map
%between observable measurements and the admissible latent
%representation.
%\end{itemize}

\medskip
\noindent
Although developed here for binary configurations, the framework extends naturally to multinomial configurations
{\small
\[
\pi_j\in\{0,1,\dots,d-1\},
\]
}

\noindent
where binary Pascal rows are replaced by multinomial simplex shells satisfying
{\small
\[
x_1+\cdots+x_d\le N.
\]
}

\noindent
For multinomial configurations, the latent representation forms a discrete simplex-supported region, while the measurement-context domain retains the corresponding multi-channel grid organization. The tensor hierarchy preserves the same principles: context-dependent aggregation, latent and context shell organization, folding-induced degeneracy, and reconstruction from structured measurements.

\medskip
\noindent
More generally, the framework suggests that observable statistical geometry may emerge naturally from combinatorial aggregation over deterministic latent ensembles.

\vspace{2mm}
\section{Interpretations}

\subsection{Dithering Interpretation}

\noindent
Earlier work on Pascal cellular automata~\cite{Richard2024} exposed an application-specific, compressed realization of the combinatorial structure considered here. Two preceding pixel values $A$ and $B$
determine a Pascal row through $|A-B|$, while $\min(A,B)$ translates the sampled or path-selected index into the admissible interval between them. The deterministic formulation additionally enumerates Pascal lattice paths through binary strings. TENSKEL recovers these local rules from the complete binary ensemble $\Omega_N=\{0,1\}^N$: the tensor hierarchy makes explicit the context-dependent multiplicities and successive foldings from which $|p-q|$ and $\min(p,q)$ emerge analytically. The earlier cellular automaton may therefore be viewed as an application-specific compressed realization of the more general combinatorial structure represented here.

\subsection{Color Interpretation}

\noindent
The multinomial extension is developed in CPrefix~\cite{Richard2026CPrefix}, where the same combinatorial principles are applied to structured discrete color mappings. The binary ensemble is replaced by a
$d$-symbol configuration ensemble and the latent Pascal triangle by the corresponding discrete multinomial simplex.

\medskip
\noindent
For a $d$-channel discretized measurement domain,
{\small
\[
\mathcal{C}^{(d)}_N
=
\{0,\ldots,N\}^{d},
\qquad
M=(N+1)^d,
\]
}
the latent representation is supported on
{\small
\[
\mathcal{I}^{(d)}_N
=
\left\{
x\in\mathcal{C}^{(d)}_N:
\|x\|_1\le N
\right\},
\qquad
K=\binom{N+d}{d}.
\]
}

\noindent
For three channels, the binary folding hierarchy becomes
{\small
\[
T(R,G,B,r,g,b)
\longrightarrow
T(R,G,B,s)
\longrightarrow
T(S,s)
\]
}
\noindent
with
{\small
\[
s=r+g+b,
\qquad
S=R+G+B.
\]
}

\noindent
CPrefix~\cite{Richard2026CPrefix} develops the corresponding latent reconstruction and evaluates it on practical color transformations, including ICC-based display and printer reconstruction and perceptual
gamut mapping in the latent simplex domain. These results provide an application-specific validation of the broader combinatorial framework developed here.

\vspace{2mm}
\section{The Story in One Breath}

\medskip
\noindent
Imagine a \emph{transparent} $(N+1)$-sided die whose visible face is the aggregated observable
\[
s\in\{0,\ldots,N\}.
\]
Inside it live two smaller dice, labelled $p$ and $q$. They are not random rolls, but measurement-context indices: two partial views into the same binary path. For a given path $\pi$, the two inner dice expose
the corresponding prefix counts
\[
a_p(\pi)
\qquad\text{and}\qquad
b_q(\pi).
\]
Their joint state $(a,b)$ belongs to the latent Pascal structure, while their sum
\[
s=a+b
\]
determines what appears on the visible surface of the transparent die.

\medskip
\noindent
If one could look through the die, the visible face would no longer appear as a single isolated outcome. For $N=5$, behind it lies the complete ensemble of $2^5=32$ binary paths. Different paths may produce
the same inner pair $(a,b)$, and different pairs may in turn fold onto the same visible value $s$. The refined tensor $\mathrm{Tab}(p,q,a,b)$ records this internal multiplicity before folding, whereas
\[
T(p,q,s)
=
\sum_{a+b=s}\mathrm{Tab}(p,q,a,b)
\]
records the resulting aggregated observation. The visible outcome is therefore only the outermost level of a structured combinatorial hierarchy:
\[
\Omega_N
\longrightarrow
(a,b)
\longrightarrow
s.
\]
Different measurement contexts partition the same deterministic latent ensemble differently, while unresolved degrees of freedom appear as combinatorial multiplicities. Once normalized, these multiplicities become the corresponding context-dependent probability distributions.

\medskip
\noindent
In this sense, the framework presented here does not begin by assigning probability to individual latent paths. It begins by counting how deterministic configurations become indistinguishable under a specified measurement context. Probability then arises by normalization of that counting structure.

\medskip
\noindent
The transparent die is therefore not proposed as a model of a physical quantum system, but as an intuitive representation of the central construction: what is observed may be a compressed view of a considerably richer combinatorial organization. 

\medskip
\noindent
\textbf{In this precise combinatorial sense, probability is the normalized shadow of counting}.

%\clearpage
\section{Conclusion}

\noindent
We introduced \textit{TENSKEL}, a combinatorial observable framework for representing how a fixed latent ensemble becomes organized under a family of measurement contexts. Starting from the complete binary
ensemble $\Omega_N=\{0,1\}^N$, the construction couples measurement contexts $(p,q)$ to latent Pascal coordinates $(a,b)$ through explicit combinatorial multiplicities. The resulting tensor therefore represents
not only which outcomes are admissible, but how measurement partitions the same latent ensemble into context-dependent observable classes.

\medskip
\noindent
This construction exposes a hierarchy that is otherwise hidden by aggregation. The refined tensor $\mathrm{Tab}(p,q,a,b)$ retains the joint context--latent structure; latent-shell folding $(a,b)\rightarrow s$ reveals embedded Pascal multiplicities; and context-shell folding $(p,q)\rightarrow S$ exposes the corresponding shell-to-shell organization. The distinction between inner and outer measurement-context regions further shows that both regions couple to the same latent Pascal support while carrying complementary information about its coordinates. In the examined binary configurations, neither context region alone provides a full-rank representation of the latent coordinates, whereas their combination restores full column rank in the complete tensor. Latent identifiability therefore emerges from the joint measurement-context organization rather than from either region in isolation.

\medskip
\noindent
The tensor representation also induces a geometry. Through $K_{\mathrm{Tab}}=\mathrm{Tab}^{\top}\mathrm{Tab}$, latent coordinates are related by the overlap of their responses across measurement contexts, while regularized inversion provides a direct reconstruction map from measurements to the latent representation. Reconstruction is therefore not introduced as an independent algorithmic principle, but arises from the observable organization encoded by the tensor itself.

\medskip
\noindent
The binary construction is not isolated. Earlier Pascal cellular automata~\cite{Richard2024} can be recovered as an application-specific compressed realization of the same combinatorial structure, while the multinomial extension developed in CPrefix~\cite{Richard2026CPrefix} replaces the binary Pascal triangle by discrete simplex-supported representations and applies the resulting tensor hierarchy to structured color measurements. These connections show how the same underlying combinatorial principle extends across dimensionalities and application domains, while the associated rank structure becomes dimension-dependent.

\medskip
\noindent
TENSKEL does not propose a new sensing modality, nor does it attempt to model a physical quantum system. Its contribution is more fundamental: it makes the organization induced by measurement an explicit mathematical object. Measurement contexts, latent configurations, observable partitions, combinatorial degeneracies, induced geometry, and reconstruction can thereby be studied within a single discrete tensor formulation. This provides a basis for further investigation of structured measurement in computational imaging, inverse problems, and structured sensing, while also establishing structural connections with measurement-centered formulations in quantum information and quantum-inspired imaging.

\medskip
\noindent
More broadly, the construction suggests a simple principle: \textbf{observable structure need not be imposed on a latent ensemble; it can emerge from the combinatorial organization induced by how that ensemble is measured.}

\vfill
\bibliographystyle{IEEEtran}
\bibliography{IEEEabrv,references}

\clearpage
\onecolumn
\appendices
\section{Hypercube Measurement Atlas}
\label{app:atlas}

\noindent
For $N=5$, the latent ensemble
\[
\Omega_5=\{0,1\}^5
\]
contains $2^5=32$ binary configurations, represented by the $32$ vertices of the $5$-dimensional hypercube. The measurement indices, however, range over the $N+1$ prefix depths
\[
p,q\in\{0,1,\ldots,5\},
\]
where the index $0$ denotes the anchored origin of the prefix construction. Equivalently, one may adjoin a fixed coordinate $\pi_0=0$ and write
\[
\widetilde{\pi}=(0,\pi_1,\ldots,\pi_5),
\]
without introducing an additional binary degree of freedom. Thus the $32$ latent configurations may be identified with the anchored facet
\[
\{0\}\times\{0,1\}^5\subset\{0,1\}^6,
\]
which is itself a $5$-dimensional hypercube. The anchor corresponds combinatorially to the zeroth Pascal row, preceding the five successive binary aggregation steps.

\medskip
\noindent
Consequently, five free binary coordinates generate six prefix depths and therefore
\[
(N+1)^2=6^2=36
\]
paired measurement contexts $(p,q)\in\{0,\ldots,5\}^2$. In particular, $(p,q)=(0,0)$ is the natural boundary context of the construction: by the empty-sum convention,
\[
a_0(\pi)=b_0(\pi)=0,
\qquad
s_{0,0}(\pi)=0
\]
for every $\pi\in\Omega_5$.

\medskip
\noindent
The atlas on the following six pages enumerates these $36$ measurement contexts on the same latent ensemble $\Omega_5$. Each panel preserves the same $32$ hypercube vertices and changes only their observable assignment
\[
s_{p,q}(\pi)=a_p(\pi)+b_q(\pi).
\]
The atlas therefore makes explicit a central feature of the construction: the latent ensemble remains fixed while the measurement context determines the observable partition induced on that ensemble. Contexts with $p=q$ collapse to a single observable class,
\[
s_{p,p}(\pi)=p,
\]
whereas increasing $|p-q|$ produces progressively wider Pascal-type multiplicity structures, consistent with the closed form derived in Section~II-D.

\medskip
\noindent
These context-dependent partitions are encoded by the observable tensor through their admissible outcomes and combinatorial multiplicities; normalization subsequently induces the corresponding probability distributions. In this limited structural sense, the atlas may also be viewed as a discrete measurement-setting representation, reminiscent of measurement-setting formulations encountered in Bell-type and contextuality scenarios, although no physical quantum interpretation is assumed here.

\medskip
\noindent
The six atlas pages are organized by fixed $p=0,\ldots,5$, with $q=0,\ldots,5$ varying across each page.

\clearpage
\includepdf[pages=-]{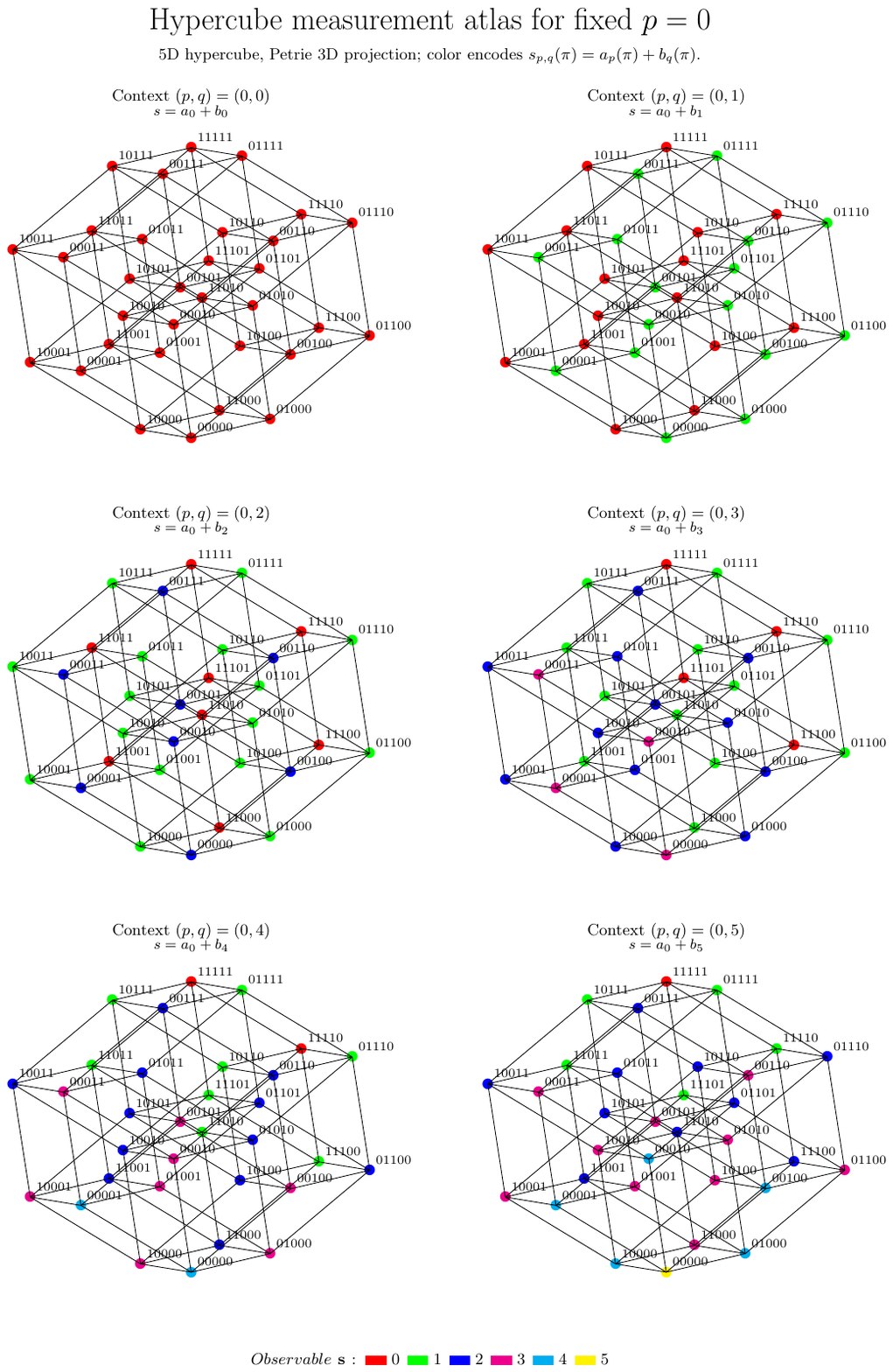}

\end{document}